\documentclass[letterpaper,11pt]{article}

\usepackage[margin=1in]{geometry}
\usepackage{setspace,graphicx,url,hyperref,natbib,enumitem}
\usepackage[ruled,linesnumbered]{algorithm2e}
\usepackage{amsmath,amsthm,amssymb,amsfonts}
\newcommand{\N}{\ensuremath{\mathbb{N}}}

\begin{document}
\title{An efficient heuristic for approximate maximum flow computations in brain connectivity networks}
\author{Jingyun Qian and Georg Hahn}
\date{}
\maketitle

\doublespacing

\begin{abstract}
Several concepts borrowed from graph theory are routinely used to better understand the inner workings of the (human) brain. To this end, a connectivity network of the brain is built, which allows one to assess quantities such as information flow and information routing via shortest path and maximum flow computations. Since brain networks typically contain thousands (or even millions) of nodes and edges, computational scaling is a key research area. In this contribution, we focus on approximate maximum flow computations in large brain networks. By combining graph partitioning with maximum flow computations, we propose a new approximation algorithm for the computation of the maximum flow with runtime $O(|V||E|^2/k^2)$ compared to the usual runtime of $O(|V||E|^2)$ for the Edmonds-Karp algorithm, where $V$ is the set of vertices, $E$ is the set of edges, and $k$ is the number of partitions. We assess both accuracy and runtime of the proposed algorithm on simulated graphs as well as on graphs downloaded from the Brain Networks Data Repository (\url{https://networkrepository.com/}).
\end{abstract}

Keywords: Approximation, Brain Connectivity Network, Edmonds Karp, Graph, Heuristic, Maximum Flow.

\section{Introduction}
\label{sec:introduction}
The human brain has been studied from a variety of angles, for instance with respect to biological, medical, and graph theoretical aspects \cite{Sporns2013,Pessoa2014}. Since the neurons and synapses in the brain form a natural network structure, a graph theoretical approach suggests itself for modeling and investigating certain properties of the brain.

Among others, information flow and information routing are two popular topics that have been studied in the literature with the help of graph theory. This includes models of information transfer in the structural brain network and their relationship to functional connectivity \cite{Neudorf2023}, new metrics to study the length of paths of cortical regions in the brain \cite{Jahanshad2012}, or resting state functional connectivity \cite{Goni2023}. Similar contributions have been made to better understand mechanisms of communication by mapping the communication problem to routing in a network \cite{Meier2015}. In this contribution, we focus on the computation of the maximum flow through a connectivity network, which is hoped to be a proxy of information flow in the brain.

The maximum flow in a capacitated and directed network from a source node to a sink node is defined as the maximum capacity that can be routed from the source to the sink via any directed path. This classic graph problem has been studied since the 1950s, with the two most prominent representatives being the Ford-Fulkerson algorithm \cite{Ford1956} and the Edmonds-Karp algorithm \cite{Edmonds1972}. The capacities on the edges must be non-negative. The interpretation of the capacities is context dependent, and in the field of brain connectivity networks the capacities usually encode some measure of information content.

Since brain connectivity networks typically contain thousands of nodes and possibly millions of edges, the computation of the maximum flow is not straightforward. This is due to the fact that the runtime of the algorithm of \cite{Edmonds1972} scales as $O(|V||E|^2)$ for an input network $N=(V,E)$, where $V$ is the set of vertices, $E$ is the set of edges, and $|\cdot|$ denotes the cardinality of a set. At the same time, we assume that the underlying brain network was measured experimentally and contains noise or errors, meaning that the maximum flow is only defined up to measurement errors in the network connectivity and the edge capacities. Moreover, it is known that brain connectivity networks often exhibit cluster-like structure \cite{Rasero2017, Cui2020}. These two observations prompt the idea to compute an approximate maximum flow by exploiting the cluster structure, in the sense that the maximum flow can be computed across clusters and then combined to yield an approximation of the exact maximum flow.

In this work, we seek to derive an algorithm with a more favorable runtime scaling than the one of \cite{Edmonds1972}, at the expense of computing an approximate maximum flow. We aim to attempt this by combining graph partitioning (using the Kernighan-Lin partitioning algorithm \cite{Kernighan1970}) with maximum flow computations. This yields an approximation algorithm with runtime $O(|V||E|^2/k^2)$ compared to the usual runtime of $O(|V||E|^2)$ for the algorithm of \cite{Edmonds1972}, where $k \in \N$ is the number of partitions. We assess both accuracy and runtime of the proposed algorithm on simulated graphs as well as graphs downloaded from the Brain Networks Data Repository \cite{brainnetworks}.

This article is structured as follows. In Section~\ref{sec:methods} we introduce the algorithms of \cite{Edmonds1972} and \cite{Kernighan1970} which we use. We then present our approximation algorithm and derive its reduced runtime compared to the original maximum flow algorithms. Section~\ref{sec:experiments} presents some experimental results to investigate accuracy and runtime of our proposed approximation heuristic, both on simulated networks and on networks downloaded from the Brain Networks Data Repository \cite{brainnetworks}. In particular, we experimentally verify the reduced runtime scaling. The article concludes with a discussion in Section~\ref{sec:discussion}.

\section{Methods}
\label{sec:methods}
This section starts with a brief review of the classical Edmonds-Karp algorithm (Section~\ref{sec:EK}), as well as the Kernighan-Lin partitioning algorithm (Section~\ref{sec:KL}), which serve as the basis of the proposed graph partitioning heuristic for the maximum flow problem. Our heuristic is presented in Section~\ref{sec:algorithm}. We conclude with a derivation of its runtime in Section~\ref{sec:runtime}.

\subsection{The Edmonds-Karp algorithm}
\label{sec:EK}
The Edmonds-Karp algorithm \cite{Edmonds1972} is a full implementation of the algorithm of Ford-Fulkerson \cite{Ford1956} for computing the maximum flow in a capacitated network. Briefly, the algorithm works by using breadth-first search (BFS) to find a so-called augmenting path from the source $s$ to the sink $t$ in $O(|E|)$ time, where for the application of BFS the edge weights are disregarded as long as they are positive. Each time a new augmenting path from $s$ to $t$ is found, that is a path from $s$ to $t$ having positive weight on all path edges, the maximal amount of flow for this path (which corresponds to the minimum of the edge capacities along the path) is pushed along the network. This is repeated until no further augmenting path exists. A detailed analysis of its runtime shows that the Edmonds-Karp algorithm runs in $O(|V||E|^2)$ time \cite{Cormen2009}.

\subsection{The Kernighan-Lin partitioning algorithm}
\label{sec:KL}
The Kernighan-Lin algorithm \cite{Kernighan1970} is a heuristic method for computing a bi-partitioning of an undirected graph with runtime $O(|V|^2\log |V|)$. Its purpose is to compute two partitions having the same or nearly the same size while minimizing the number of cut edges between them. In our application, we focus on the number of cut edges only, thus disregarding the edge weights. To compute more than two partitions, we implement the Kernighan-Lin algorithm recursively to achieve a $2^l$-partitioning, where $l \in \N$ represents the level or the iterations of the bi-partitioning.

\subsection{A heuristic for computing the maximum flow}
\label{sec:algorithm}
\begin{figure}
    \centering
    \includegraphics[width=0.4\textwidth]{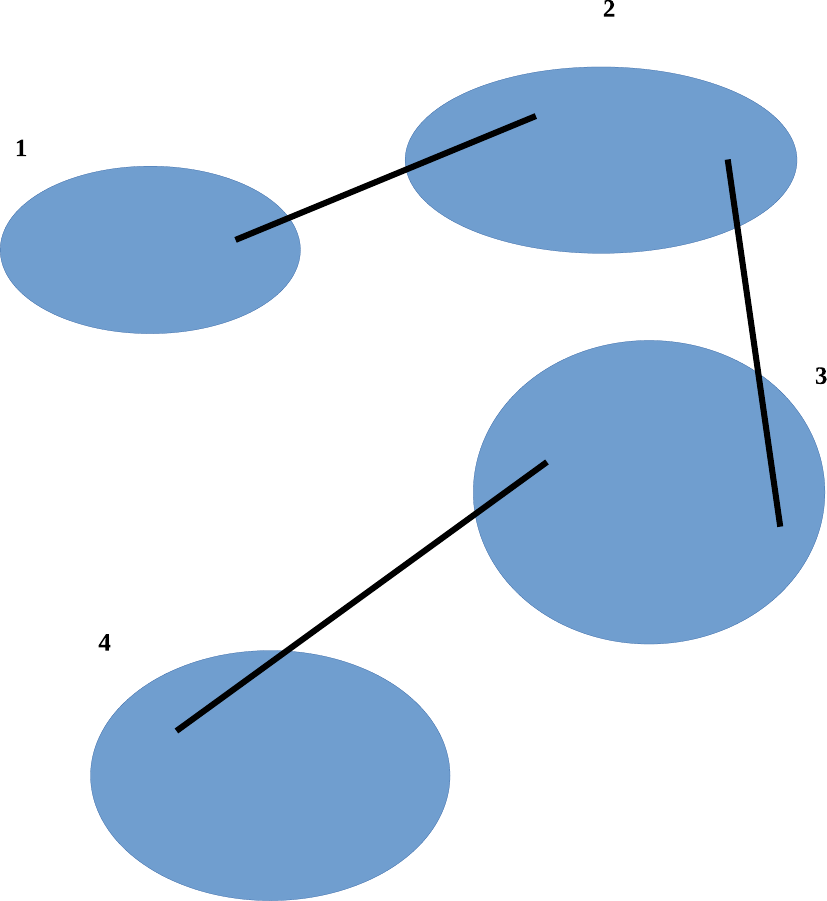}
    \caption{Schematic idea of the computation of the maximum flow between and within partitions.}
    \label{fig:schematic}
\end{figure}

Our heuristic algorithm strategically partitions the graph into $k$ parts for $k \in \N$ to speed up the computation of the maximum flow for each subgraph. The idea of the heuristic is visualized in Figure~\ref{fig:schematic}. It is based on the fact that a heuristic partitioning of a graph can be computed cheaply, and that the computation of the flow for the partitions separately is computationally cheaper than on the entire graph. We thus first partition the graph into $k$ partitions using a recursive application of the Kernighan-Lin algorithm. The choice of $k \in \N$ is arbitrary and will be investigated in Section~\ref{sec:experiments}.

After partitioning, two scenarios are possible for computing the flow between some source node $s$ and sink node $t$. If $s$ and $t$ reside in the same cluster after partitioning, the maximum flow within their partition from $s$ to $t$ is assumed to be a good approximation of the actual maximum flow. On the other hand, if $s$ and $t$ are in separate clusters, we treat each cluster subgraph as a ``supernode" and approximate the maximum flow within each cluster as the flow between two randomly chosen vertices. Other choices could involve using distance metrics to more judiciously select the nodes that connect the clusters.

After the flow in each cluster is computed, we conceptualize the network $N$ as consisting of $k$ supernodes, with each supernode being one cluster. Two supernodes are connected with an edge of infinite capacity if their corresponding clusters are connected with least one edge. This directed connection matrix allows our approach to adapt to various graph types. The overall maximum flow is then determined by identifying all (shortest) paths between the two supernodes containing $s$ and $t$. The final flow approximation is computed as the sum of all flows which can be pushed from the supernode containing $s$ to the supernode containing $t$, while respecting the constraints that the flow through a supernode (which is a cluster) cannot exceed the previously computed maximum flow in that cluster.

\begin{algorithm}[t]
    \caption{Heuristic Maximum Flow Algorithm}
    \label{algo:approximation}
    \SetAlgoLined
    \SetKwInOut{Input}{input}
    \Input{network $N=(V,E)$, source $s \in V$, sink $t \in V$, number of partitions $k \in \N$\;}
    Compute $k$ partitions $C_1,\ldots,C_k$ of $N$ with Kernighan-Lin\;
    Let $i,j \in \{1,\ldots,k\}$ denote the indices of the partitions containing $s$ and $t$, respectively\;
    \If{$C_i = C_j$}{
        Extract subgraph $G_i$ induced by $C_i$\;
        Compute maximum flow $f$ in $G_i$ from $s$ to $t$\;
        \Return $f$\;
    }
    \Else{Compute maximum flow values $f_r$ for partition $C_r$ for $r \in \{1,\ldots,k\}$, where for $f_i$ the flow is from the source $s$ to a random node, for $f_j$ the flow is from a random node to the sink $t$, and for all others between random nodes\;
    $G' \leftarrow$ graph induced by the partitions $C_1,\dots,C_k$ and edges between them\;
    $f_{G'} \leftarrow 0$\;
    \While{$\exists$ path in $G'$ from $C_i$ to $C_j$}{
        Find the shortest path $p$ from $C_i$ to $C_j$\;
        Let $\mu \leftarrow \min_{r:C_r \in p} f_r$ be the minimal flow value along $p$\;
        $f_{G'} \leftarrow f_{G'} + \mu$\;
        Reduce flow along $p$, that is $f_r \leftarrow f_r - \mu$ for all $r$ such that $C_r \in p$\;
        If $f_r=0$ then remove $C_r$ for all $r \in \{1,\ldots,k\}$\;
    }
    \Return $f_{G'}$}
\end{algorithm}

Our proposed heuristic is summarized in Algorithm~\ref{algo:approximation}. Its input are a network $N=(V,E)$ for which the maximum flow is sought between some input source $s \in V$ and sink $t \in V$, and the number of partitions $k \in \N$ used for the approximation.

Algorithm~\ref{algo:approximation} starts by running the Kernighan-Lin algorithm to compute $k$ partitions $C_1,\ldots,C_k$ of the network $N$. Moreover, the indices $i,j \in \{1,\ldots,k\}$ of the partitions containing $s$ and $t$ are determined. Then, two cases are being distinguished. First, if $s$ and $t$ are in the same partition $C_i=C_j$, we approximate the maximum flow by computing it within $C_i$ only. Second, we consider the case where $s$ and $t$ are not located within the same partition. In this case, we compute the maximum flow $f_r$ in each partition $C_r$, where $r \in \{1,\ldots,k\}$. Note that for $C_i$ containing the source $s \in V$, we compute the maximum flow between $s$ and a random node, which serves as a proxy for the node connecting the partition to another partition in a flow computation. Similarly, for $C_j$ containing the sink $t$, we compute the maximum flow between a random node and $t$. For all other partitions, the maximum flow is computed between two random nodes. Finally, we need to assemble the different flows into one approximate flow between $s$ and $t$. This is done by creating a new graph $G'$ in which each cluster (partition) $C_i$ becomes a supernode, and supernodes are connected by an edge of infinite capacity if there exists at least one edge between the corresponding partitions. Note that $G'$ has $k$ vertices and at most $k^2$ edges, so any computation on $G'$ is independent of $|V|$ and $|E|$. Then, we compute all paths between $C_i$ and $C_j$ and determine the minimal amount $\mu$ of flow that can be pushed along each path, which is restricted by the amounts that can flow through the partitions $C_r$, $r \in \{1,\ldots,k\}$. For each such path, the amount $\mu$ is added to some variable $f_{G'}$ initialized with zero, and after all paths with positive flow are explored, the algorithm terminates and returns $f_{G'}$ as our maximum flow approximation.

\subsection{Runtime}
\label{sec:runtime}
The runtime of the heuristic is determined by four contributions, precise the partitioning step, the computation of the maximum flow in all $k$ partitions, the building of the connectivity graph $G'$ in which each partition is a supernode, and the path computations in $G'$. In this section, we will discuss all four runtime contributions.

First, the runtime of one bi-partitioning step of the Kernighan-Lin algorithm is $O(|V|^2\log |V|)$, where $V$ is the vertex set to be partitioned \cite{Kernighan1970}. We call the Kernighan-Lin algorithm in a recursive fashion for $l \in \N$ bisection levels, resulting in $k=2^l$ partitions. During the bisectioning process, on each level $i \in \{0,\ldots,l\}$ the graph already contains $2^i$ partitions, and we call Kernighan-Lin again on each of those (meaning $2^i$ times) to further split these partitions. Since each partition on level $i \in \{0,\ldots,l\}$ contains $|V|/2^i$ vertices, the runtime can be bounded by
\begin{align*}
    \sum_{i=0}^l 2^i \left( \left( \frac{|V|}{2^i} \right)^2 \log \left( \frac{|V|}{2^i} \right) \right) &\leq \sum_{i=0}^l 2^i \left( \left( \frac{|V|}{2^i} \right)^2 \log |V| \right)\\
    &\leq |V|^2\log |V| \sum_{i=0}^\infty \frac{1}{2^i}.
\end{align*}
As the geometric series $\sum_{i=0}^\infty 2^{-i}=2$ converges to a constant, the runtime remains of the order of $O(|V|^2\log |V|)$.

Second, running the Edmonds-Karp algorithm on the $k$ partitions, each having $|V|/k$ nodes and $|E|/k$ edges, results in a runtime of $O \left( \frac{|V|}{k} \frac{|E|^2}{k^2} \right)$ for each partition. Since we run it for all $k$ partitions, the overall runtime is accordingly $O \left( \frac{|V| |E|^2}{k^2} \right)$.

Third, the connectivity network $G'$ in which each partition is a supernode is constructed in $O(k^2)$, which is also the size of its adjacency matrix.

Fourth, the path computation in $G'$ can be accomplished with the help of, for instance, the Floyd-Warshall algorithm \cite{Floyd1962,Warshall1962} in runtime $O(k^3)$.

Together, the runtime of the entire algorithm is bounded by $O \left( \frac{|V| |E|^2}{k^2} + |V|^2 \log |V| + k^3 \right)$. For dense graphs, meaning those for which $|E| \simeq |V|^2$, the dominating term in the runtime is $\frac{|V| |E|^2}{k^2} \simeq \frac{|V|^5}{k^2}$. Therefore, one expects a considerable runtime improvement for networks that exhibit a cluster structure with a large number of clusters (partitions) $k$. Even for sparse graphs with $|E| \simeq |V|$, the dominating term in the runtime is still $\frac{|V| |E|^2}{k^2} \simeq \frac{|V|^3}{k^2}$, thus for suitable values of $k$ a considerable runtime improvement is expected.

\section{Experimental results}
\label{sec:experiments}
We start by describing the experimental setting (Section~\ref{sec:setting}), followed by a description of the Brain Networks Data Repository whose benchmark graphs we use (Section~\ref{sec:database}). In Section~\ref{sec:simulated}, we present experimental evidence demonstrating that the runtime of our proposed heuristic scales favorably and as theoretically predicted while maintaining a comparable level of accuracy with respect to the traditional Edmonds-Karp algorithm. These experiments are conducted on sparse and clustered graphs, as well as sparse, non-clustered graphs. In Section~\ref{sec:brain}, we apply our heuristic algorithm to graphs downloaded from the Brain Networks Data Repository.

\subsection{Setting}
\label{sec:setting}
\begin{figure}
    \centering
    \includegraphics[width=0.4\textwidth]{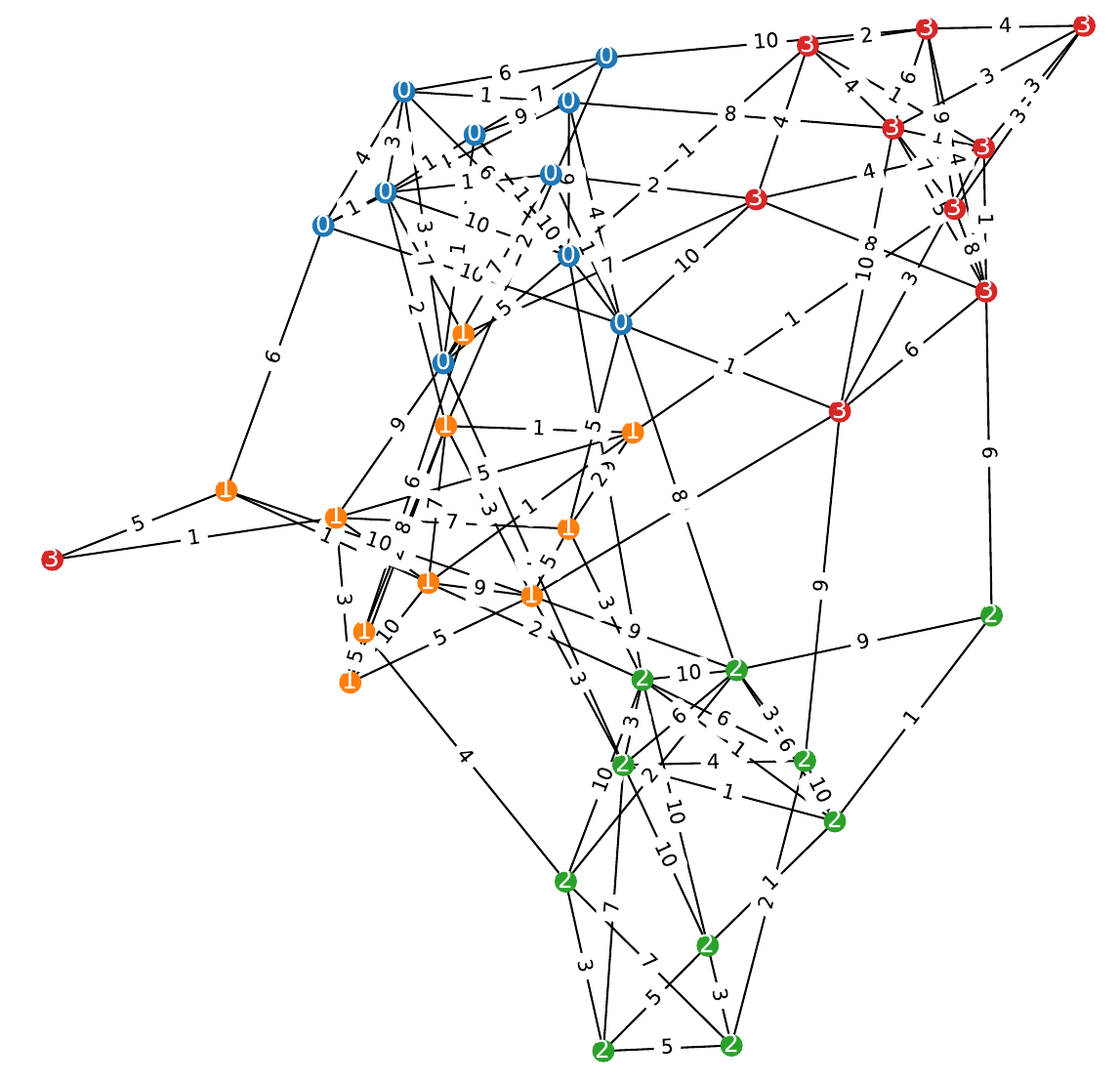}
    \caption{An example of a clustered graph with 4 clusters and 10 nodes per cluster, generated with parameters \emph{inp} = 0.5, and \emph{outp} = 0.05. The vertex coloring reflects the clusters identified by the Kernighan-Lin algorithm.}
    \label{fig:example}
\end{figure}

To simulate clustered graphs at random, we use the NetworkX function \textit{random\_partition\_graph} and assign edge weights drawn uniformly at random from the integer set $\{1,\ldots,10\}$. All code was implemented on an Apple M2 Pro Chip running macOS Ventura 13.4, using Python 3.11.5 compiled with Clang 14.0.6. Our simulation setup has four parameters: the number of clusters, the number of nodes per cluster, and the probabilities of edge formation within (\emph{inp}) and between (\emph{outp}) clusters. For example, in Figure~\ref{fig:example}, we generate a graph with 4 clusters, each containing 10 nodes, with \emph{inp} = 0.5 and \emph{outp} = 0.05. The coloring of the graph is chosen to depict the clusters identified by the Kernighan-Lin partitioning algorithm.

\subsection{The Brain Networks Data Repository}
\label{sec:database}
The Brain Networks Data Repository \cite{brainnetworks} is the largest real-time interactive repository with thousands of network graphs in more than 30 domains, ranging from biological to social networks. The brain network database aims to provide benchmark networks to facilitate scientific studies in machine learning and network science. Specifically, we utilize brain network data organized into adjacency lists, covering a variety of structures including the Drosophila medulla, mouse retina and brain, Macaque-rhesus brain, as well as cortical and cerebral regions \cite{bigbrain}. The references to each dataset can be found in Section~\ref{sec:availability}.

\subsection{Results on clustered and non-clustered graphs}
\label{sec:simulated}
We start by investigating the scaling of Algorithm~\ref{algo:approximation} and measure both the runtime and the approximated maximum flow value (compared with the exact value obtained with the classic Edmonds-Karp algorithm). To this end, we apply Algorithm~\ref{algo:approximation} to randomly generated graphs with $k=2^3$ partitions, where we use a within-cluster edge probability of \emph{inp} = 0.01, and a between-cluster edge probability of \emph{outp} = 0.005. The number of nodes is varied from $10^2$ to $10^5$.

\begin{figure*}
    \centering
    \includegraphics[width=\textwidth]{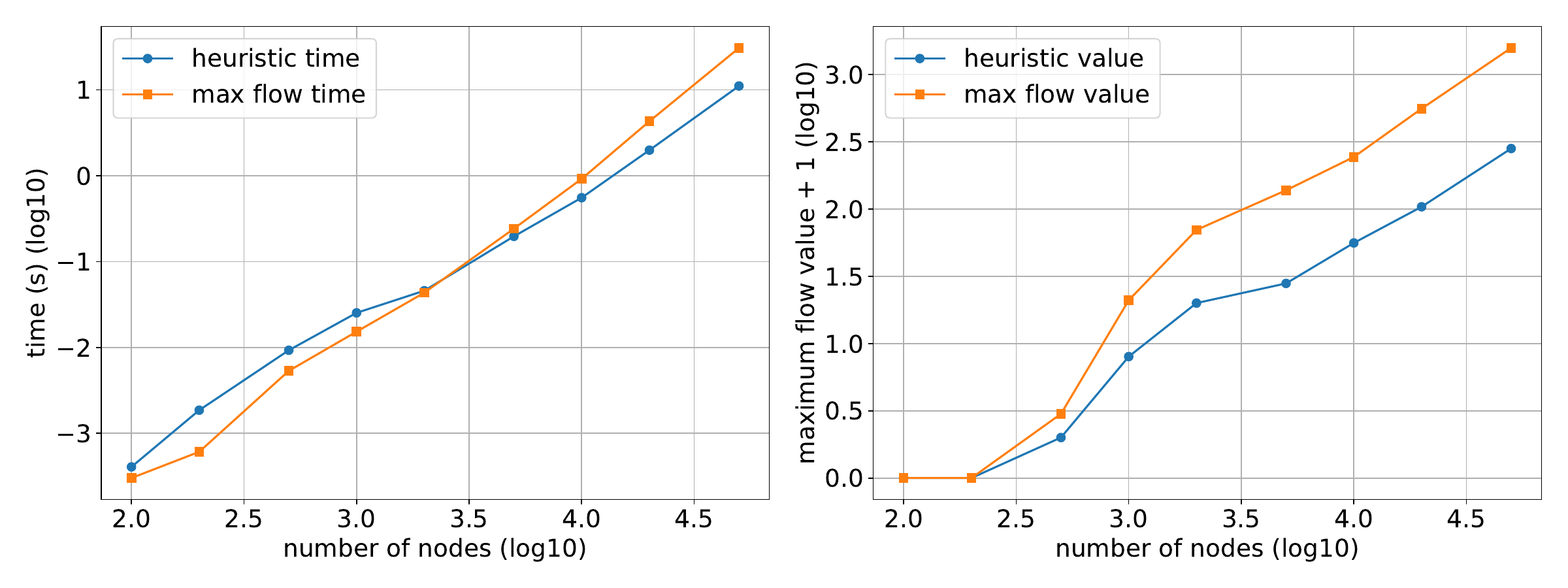}
    \caption{Runtime (left) and maximum flow value (right) in sparse and clustered graphs as a function of the number of nodes, where \emph{inp} = 0.01 and \emph{outp} = 0.005. The number of assumed clusters is $k=2^3$. Algorithm~\ref{algo:approximation} (heuristic) in blue and exact value obtained with Edmonds-Karp in orange. Log scale on both axes.}
    \label{fig:sparse_clustered}
\end{figure*}

Figure~\ref{fig:sparse_clustered} shows our results. We find that the runtime of Algorithm~\ref{algo:approximation} seems to be comparable to the one of the classic Edmonds-Karp algorithm for a small number of nodes, though Algorithm~\ref{algo:approximation} seems to be faster as the number of nodes increases. The runtimes for both Algorithm~\ref{algo:approximation} and Edmonds-Karp exhibit a linear trend in the log-log plot, thus indicating a polynomial runtime scaling which is expected. The slope for Algorithm~\ref{algo:approximation} is lower, indicating a more favorable runtime scaling. At the same time, our heuristic seems to systematically underestimate the maximum flow value, though the overall trend is the same as for the classic Edmonds-Karp algorithm.

\begin{figure*}
    \centering
    \includegraphics[width=1\textwidth]{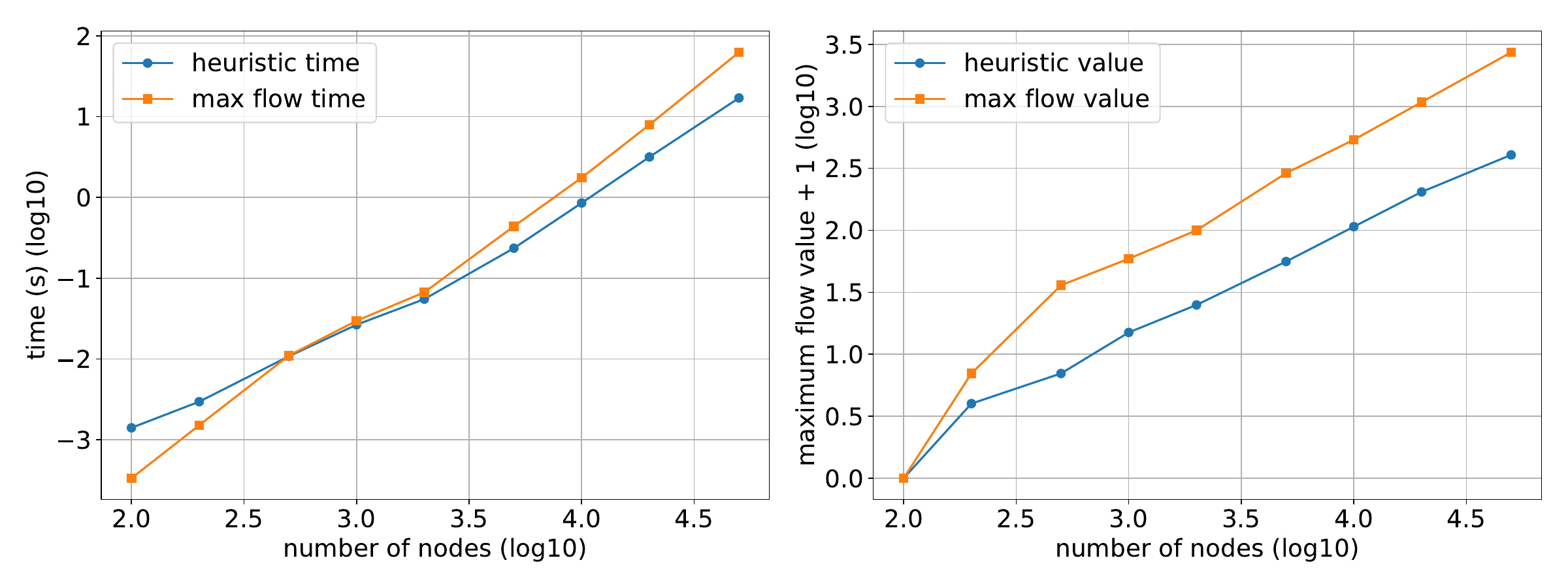}
    \caption{Runtime (left) and log maximum flow value (right) in sparse and non-clustered graphs as a function of the number of nodes, where \emph{inp} = 0.01 and \emph{outp} = 0.01. The number of assumed clusters is $k=2^3$. Algorithm~\ref{algo:approximation} (heuristic) in blue and exact value obtained with Edmonds-Karp in orange. Log scale on both axes.}
    \label{fig:sparse_notclustered}
\end{figure*}

Second, we repeat the same experiment for sparse and non-clustered graphs in Figure~\ref{fig:sparse_notclustered}. Here, the picture is similar despite the fact that the graphs are generated with both a within-cluster and between-cluster probability of \emph{inp} = \emph{outp} = 0.01, meaning without cluster structure. Nevertheless, Algorithm~\ref{algo:approximation} is run with $k=2^3$ assumed clusters. Interestingly, the returned maximum flow value does not seem to be affected by much.

Third, we aim to empirically verify the theoretical runtime of Algorithm~\ref{algo:approximation}, derived as $|V| |E|^2 / k^2$ in Section~\ref{sec:runtime}. In particular, we aim to verify the dependence of $k^{-2}$ on the number of partitions $k \in \N$. To this end, we fix the between-cluster probability \emph{outp} = 0.005 and the within-cluster probability \emph{inp} = 0.01, and generate random graphs with a varying number of nodes from 100 to 1000 in steps of 200. We record the runtime of Algorithm~\ref{algo:approximation} as a function of $k=2^l$, where $l \in \{0,2,4,6,8,10\}$. All results are averages over 100 repetitions. To demonstrate the theoretical runtime dependence of $|V| |E|^2 / k^2$ on the number of partitions $k=2^l$, we compute the ratio of the empirically measured runtimes for any $k=2^l$ with $l>0$ to the case $l=0$ (that is, to the empirical runtime for $k=2^0=1$ partition), thus expecting a $2\log(k)$-fold decrease in runtime on the log scale.

\begin{figure*}
    \centering
    \includegraphics[width=1\textwidth]{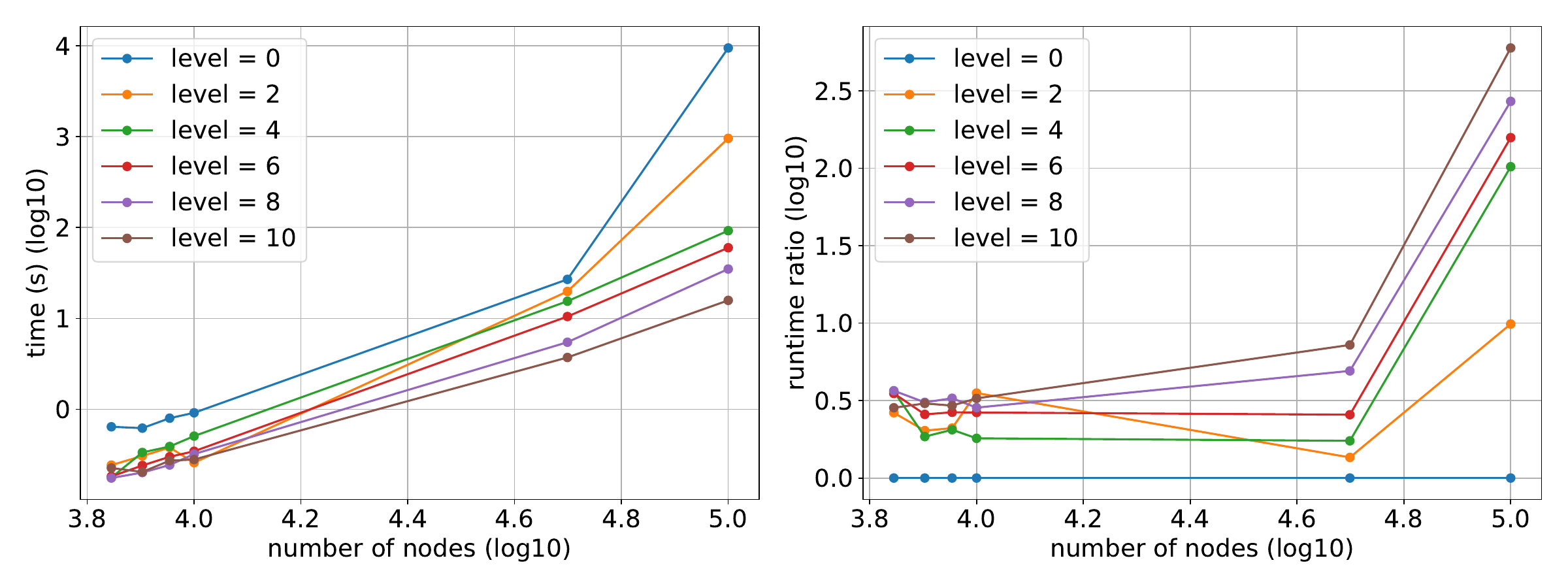}
    \caption{Runtime (left) as a function of the number of partitions $2^l$, where the level $l=\{0,2,4,6,8,10\}$ is given in the legend. The ratio (right) of the runtime for any level $l>0$ to the one at $l=0$. Log scale on both axes. Note that the runtime ratio for $l=0$ to itself is one, displayed as $\log(1)=0$ in the figure.}
    \label{fig:levels}
\end{figure*}

Our results are shown in Figure~\ref{fig:levels}. Two observations are noteworthy. First, Figure~\ref{fig:levels} (left) demonstrates that as expected, the runtimes decrease as $k$ increases. Second, Figure~\ref{fig:levels} (right) shows the log ratios of the runtimes for $l>0$ to the baseline case of $l=0$ with $k=2^0=1$ partition. We observe that indeed, for the most part, the ratios are constant apart from the case of a large number of nodes. However, the value of the ratios differs from the predicted value of $2\log(k)$, a fact that is attributed to measurement error.

\subsection{Application to brain network graphs}
\label{sec:brain}
To test Algorithm~\ref{algo:approximation} on real-world data, we apply it to $6$ different datasets with at most 2000 nodes. To avoid obtaining a zero maximum flow value (which occurs for any pair of nodes that is not connected by a path), we randomly selected source and sink nodes from the top 20 nodes with the highest degree.

\begin{table*}
    \centering
    \begin{tabular}{l|rr|rr|rr}
        \hline
        Dataset & Nodes & Edges & Time (Alg.~\ref{algo:approximation}) & Time (EK) & Value (Alg.~\ref{algo:approximation}) & Value (EK)\\
        \hline
        Drosophila Medulla & 1781 & 9735 & \textbf{0.047} & 0.054 & 145 & 145 \\
        Mouse Retina & 1076 & 90811 & 0.336 & \textbf{0.251} & 205 & 307 \\
        Macaque-rhesus Brain & 242 & 4090 & \textbf{0.006} & 0.012 & 25 & 30 \\
        Mouse Brain & 213 & 21807 & \textbf{0.022} & 0.085 & 52 & 84 \\
        Macaque-rhesus Cortical & 93 & 2667 & \textbf{0.002} & 0.006 & 13 & 28 \\
        Macaque-rhesus Cerebral & 91 & 1615 & 0.006 & 0.006 & 11 & 20 \\
        \hline
    \end{tabular}
    \caption{Performance of Algorithm~\ref{algo:approximation} and Edmonds-Karp (EK) on graphs downloaded from the Brain Network Data Repository \cite{brainnetworks}.}
    \label{tab:experiment}
\end{table*}

Results are presented in Table~\ref{tab:experiment}, showing the graph characteristics (dataset name, number of nodes and edges), the runtime of Algorithm~\ref{algo:approximation} and the obtained maximum flow value, as well as the runtime and maximum flow value for the exact computation with the Edmonds-Karp algorithm. The fastest runtimes are highlighted in bold font. The table illustrates that Algorithm~\ref{algo:approximation} is generally faster than the classic Edmonds-Karp algorithm, yet Algorithm~\ref{algo:approximation} is able to closely approximate the exact maximum flow values in real-world data applications.

\section{Discussion}
\label{sec:discussion}
This article presents a new approximation algorithm to compute the maximum flow of a capacitated network. Our algorithm is designed for clustered graphs. It is based on the observation that a heuristic partitioning of a graph can be computed cheaply, and that the computation of the maximum flow for each partition separately is computationally cheaper than on the entire graph.

We demonstrate theoretically that the runtime of our algorithm is proportional to $|V| |E|^2 / k^2$, where $V$ is the vertex set, $E$ is the edge set, and $k$ is the number of partitions, thus considerably reducing the observed runtimes for graphs that can be clustered into a large number of partitions $k$.

In an experimental section, we showcase both accuracy and runtime of our heuristic using simulated networks and networks downloaded from the Brain Network Data Repository. Our empirical results confirm that the runtime indeed decreases as the number of partitions $k$ increases, and that on real-world graphs, our heuristic is able to compute approximate maximum flow values of high quality while having a faster runtime.

\subsection*{Acknowledgements}
The authors thankfully acknowledge the contributors of the Brain Networks Data Repository.

\appendix
\section{Availability of datasets}
\label{sec:availability}
The datasets analyzed in this study are available from the Brain Networks Data Repository, see \url{https://networkrepository.com/}. They have the following URL identifiers:

\begin{enumerate}[leftmargin=*,noitemsep]
    \item \url{https://networkrepository.com/bn-fly-drosophila-medulla-1.php}
    \item \url{https://networkrepository.com/bn-mouse-retina-1.php}
    \item \url{https://networkrepository.com/bn-macaque-rhesus-brain-1.php}
    \item \url{https://networkrepository.com/bn-mouse-brain-1.php}
    \item \url{https://networkrepository.com/bn-macaque-rhesus-interareal-cortical-network-2.php}
    \item \url{https://networkrepository.com/bn-macaque-rhesus-cerebral-cortex-1.php}
\end{enumerate}

\bibliographystyle{plain}
\bibliography{main}
\end{document}